\documentclass[reprint,bibnotes,superscriptaddress,amsmath,amssymb,longbibliography,aps,prx]{revtex4-2}

\usepackage{graphicx}
\usepackage{amsmath}
\usepackage{amssymb}
\usepackage{color}
\usepackage{comment}
\usepackage{bm}
\usepackage[colorlinks=true,linkcolor=blue,citecolor=blue]{hyperref}
\usepackage{xcolor}

\begin{document}

\title{High-field fate of the Kitaev quantum spin liquid in $\alpha$-RuCl$_3$}

\author{K.~Imamura}
\email{imamura.kumpei.3h@kyoto-u.ac.jp}
\altaffiliation{Present address: Department of Physics, Kyoto University, Sakyo-ku, Kyoto 606-8502, Japan}
\affiliation{Department of Advanced Materials Science, University of Tokyo, Kashiwa, Chiba 277-8561, Japan}
\author{R.~Ohno}
\affiliation{Department of Physics, Tohoku University, Sendai 980-8578, Japan}
\author{Y.~C.~Tsuzuki}
\affiliation{Department of Advanced Materials Science, University of Tokyo, Kashiwa, Chiba 277-8561, Japan}
\author{Y.~Akui}
\affiliation{Department of Advanced Materials Science, University of Tokyo, Kashiwa, Chiba 277-8561, Japan}
\author{R.~Namba}
\affiliation{Department of Advanced Materials Science, University of Tokyo, Kashiwa, Chiba 277-8561, Japan}
\author{K.~Ishihara}
\affiliation{Department of Advanced Materials Science, University of Tokyo, Kashiwa, Chiba 277-8561, Japan}
\author{M.~Akaki}
\affiliation{Institute for Materials Research, Tohoku University, Sendai 980-8577, Japan}
\author{M.~Kimata}
\affiliation{Institute for Materials Research, Tohoku University, Sendai 980-8577, Japan}
\affiliation{Advanced Science Research Center, Japan Atomic Energy Agency, Tokai, Ibaraki 319-1195, Japan}
\author{N.~Kurita}
\affiliation{Department of Physics, Institute of Science Tokyo, Meguro, Tokyo 152-8551, Japan}
\author{H.~Tanaka}
\affiliation{Center for Entrepreneurship Education, Institute of Science Tokyo, Yokohama 226-8502, Japan}
\author{N.~Kimura}
\affiliation{Department of Physics, Tohoku University, Sendai 980-8578, Japan}
\author{S.~Imajo}
\affiliation{Department of Advanced Materials Science, University of Tokyo, Kashiwa, Chiba 277-8561, Japan}
\affiliation{Institute for Solid State Physics, The University of Tokyo, Kashiwa, Chiba 277-8581, Japan}
\author{A.~Matsuo}
\affiliation{Institute for Solid State Physics, The University of Tokyo, Kashiwa, Chiba 277-8581, Japan}
\author{K.~Kindo}
\affiliation{Institute for Solid State Physics, The University of Tokyo, Kashiwa, Chiba 277-8581, Japan}
\author{Y.~Matsuda}
\affiliation{Department of Physics, Kyoto University, Sakyo-ku, Kyoto 606-8502, Japan}
\affiliation{Los Alamos National Laboratory, Los Alamos, New Mexico 87545, USA}
\author{K.~Hashimoto}
\affiliation{Department of Advanced Materials Science, University  of Tokyo, Kashiwa, Chiba 277-8561, Japan}
\affiliation{Department of Physics, Kyoto University, Sakyo-ku, Kyoto 606-8502, Japan}
\author{Y.~Mizukami}
\affiliation{Department of Physics, Tohoku University, Sendai 980-8578, Japan}
\author{T.~Shibauchi}
\email{shibauchi@k.u-tokyo.ac.jp}
\affiliation{Department of Advanced Materials Science, University of Tokyo, Kashiwa, Chiba 277-8561, Japan}

\begin{abstract}
Kitaev quantum spin liquids (KQSLs) host fractionalized excitations described by itinerant Majorana quasiparticles and gapped $Z_2$ fluxes (visons), providing a platform for emergent topological matter. 
Whether such a state survives under strong magnetic fields, however, remains an open question. The layered honeycomb magnet $\alpha$-RuCl$_3$ is a leading candidate material: an in-plane field of $\sim 7$\,T suppresses antiferromagnetic order and induces a quantum-disordered phase exhibiting signatures consistent with Majorana excitations, including an anomalous thermal Hall effect and field-angle-dependent specific heat. At higher fields, the magnetization approaches saturation, suggesting a transition to a spin-polarized state, yet the microscopic evolution between these limits remains unresolved. 
Here we report high-field specific heat measurements up to 24\,T that reveal a distinct crossover at $\mu_0H^*\approx 15$\,T, beyond which the perturbative Kitaev description breaks down. Above $H^*$, the characteristic six-fold angular modulation of the specific heat collapses and the excitation gap deviates from the predicted $H^3$ scaling. 
Meanwhile, the gap decreases with increasing field and the in-plane magnetization anisotropy persists up to $\sim24$\,T, both in sharp contrast to a trivial spin-polarized state, indicating that KQSL signatures are preserved even at $\sim 90$\,\% of magnetization saturation. These results reveal that the KQSL in $\alpha$-RuCl$_3$ extends well beyond the perturbative window, persisting as a nonperturbative regime in which the Majorana and vison energy scales merge, before eventually giving way to spin polarization.
This thermodynamic roadmap provides a basis for understanding how fractionalized phases evolve under strong magnetic fields.

\end{abstract}

\maketitle

\section{Introduction}

Quantum spin liquids are exotic quantum phases of matter in which strong quantum fluctuations preclude the formation of conventional symmetry-breaking magnetic order, and instead give rise to unconventional quasiparticles associated with emergent gauge fields and topological order~\cite{RevModPhys.89.025003}. 
A central theoretical reference is the Kitaev honeycomb model, which is exactly solvable and realizes KQSLs.
In the KQSL, the excitations admit an explicit quasiparticle description in terms of itinerant Majorana fermions and gapped $Z_2$ fluxes~\cite{KITAEV20062}.
In zero field, the itinerant Majorana spectrum is gapless featuring Dirac-like cones, which provides a well-defined low-energy structure separated from the gapped $Z_2$ flux energy scale.
A magnetic field breaks time-reversal symmetry and, in the perturbative regime, generates an effective three-spin term $\kappa \propto h_xh_yh_z/\Delta_{\text{flux}}^2$.
Here $h_i$ with $i=x,y,z$ are the components of the applied magnetic field $\bm{H}$, and $\Delta_{\text{flux}}$ is the excitation gap of $Z_2$ flux, which is proportional to Kitaev exchange scale $K$.
This three-spin term opens a Majorana gap $\Delta_{\text M}$ in the low-energy spectrum, with $\Delta_{\text M}\propto|\kappa|$. 
The resulting gapped Majorana bands are topological, characterized by a Chern number $Ch=\pm1$ with ${\text{sgn}}(Ch)={\text{sgn}}(h_xh_yh_z)$. 
Through bulk-edge correspondence, a nonzero $Ch$ leads to a chiral Majorana edge mode~\cite{KITAEV20062}.

Motivated by this theoretical framework, the layered honeycomb magnet $\alpha$-RuCl$_3$ has been intensively studied as a leading candidate material proximate to a KQSL~\cite{Matsuda2025,takagi2019concept}.
Despite a dominant ferromagnetic Kitaev exchange with a coupling strength $K\approx$ 5-10\,meV~\cite{suzuki2021proximate,PhysRevResearch.2.033011}, non-Kitaev interactions, including Heisenberg and off-diagonal terms, induce a zigzag (ZZ) antiferromagnetic (AFM) order below $T_{\text N}\approx 7$\,K~\cite{PhysRevB.92.235119}.
Even so, the paramagnetic regime above $T_{\text N}$ exhibits multiple signatures consistent with Kitaev-like spin fractionalization as revealed by neutron and Raman scattering and by thermodynamic measurements~\cite{banerjee2016proximate,banerjee2017neutron,PhysRevLett.114.147201,nasu2016fermionic,do2017majorana,PhysRevB.99.094415}.
Upon applying an in-plane magnetic field, the zigzag AFM phase is suppressed above $\approx7$ T and the system enters the field-induced quantum-disordered (FIQD) state~\cite{yadav2016kitaev,PhysRevB.96.041405,banerjee2018excitations}.
Most notably, a half-integer quantized thermal Hall response has been reported and interpreted in terms of chiral Majorana edge transport expected for a topological KQSL~\cite{kasahara2018majorana,yokoi2020half,PhysRevB.106.L060410,PhysRevB.102.220404,bruin2022robustness,xing2025magnetothermal}, although its definitive verification remains under active scrutiny~\cite{Ongplanar,lefranccois2022evidence,Zhang2024}.
From a bulk thermodynamic perspective, in-plane field-angle-resolved specific heat measurements provide a sensitive bulk probe of charge-neutral Majorana excitations because the low-energy spectrum can acquire a strongly field-angle-dependent Majorana gap $\Delta_{\text M}\propto|\kappa|$ in the perturbative field regime~\cite{tanaka2020thermodynamic,imamura2024majorana}.
In the FIQD regime of $\alpha$-RuCl$_3$, the excitation gap shows a pronounced in-plane anisotropy~\cite{tanaka2020thermodynamic}: it is largest for fields along the zigzag directions ($\bm{H}\parallel\bm{a}$) and closes for fields along the Ru–Ru bond (armchair) directions ($\bm{H}\parallel\bm{b}$), yielding gapless Dirac-cone excitations~\cite{imamura2024majorana}, a hallmark of the topological nature.
Together with the reported in-plane angular dependence of thermal Hall signals, this bulk thermodynamic anisotropy has been discussed in terms of Majorana fermion excitations and bulk–edge correspondence in a proximate KQSL state.
Recent progress in crystal growth, most notably the two-step sublimation method, now yields substantially higher-quality $\alpha$-RuCl$_3$ crystals~\cite{namba2024}. 
Crucially, the same anisotropic Majorana excitations are fully reproduced in these ultraclean samples~\cite{PhysRevB.111.184440}, supporting that the KQSL state is intrinsic to the FIQD regime of $\alpha$-RuCl$_3$ and suggesting, from a bulk perspective, robustness of the essential signatures against moderate disorder~\cite{PhysRevB.111.184440,Imamura2024defect}.
More recently, thermal Hall measurements on microscale samples have provided further direct evidence for chiral fermionic edge transport~\cite{HedaZhang2025}.

Building on the preceding discussion, the behavior expected in the perturbative KQSL regime is now well established in $\alpha$-RuCl$_3$ by several experiments. 
However, these experimental studies of Majorana excitations in $\alpha$-RuCl$_3$ have been mainly focused on the laboratory field range (up to 12-14\,T), leaving the fate of the perturbative Kitaev description at higher fields largely unexplored. 
In particular, the field range over which the characteristic KQSL signatures survive has not been established.
Starting from a KQSL, increasing the field must ultimately drive the system toward the spin-polarized (SP) limit, but theoretically, the route to this limit depends strongly on whether the underlying Kitaev exchange is AFM or ferromagnetic (FM)~\cite{janssen2019heisenberg,feng2026magnetic,hickey2019emergence,PhysRevB.100.165123,patel2019magnetic,zhang2022theory,lee2020magnetic,li2021identification,gordon2019theory}.
For the AFM Kitaev model, increasing the field can generate nontrivial intermediate quantum phases between the KQSL and SP phases, including gapless $U(1)$ spin-liquid regimes and field-induced states with distinct topological character, such as different Chern numbers~\cite{hickey2019emergence,PhysRevB.100.165123,patel2019magnetic,zhang2022theory}.
In contrast, for the FM Kitaev model for an in-plane magnetic field, many theoretical calculations find a comparatively direct evolution from the KQSL to the SP phase by a single transition~\cite{hickey2019emergence,zhang2022theory,lee2020magnetic,li2021identification}.
Experimentally, high-field magnetization measurements extending beyond 50\,T indicate that the magnetization approaches saturation at sufficiently large in-plane fields, implying that $\alpha$-RuCl$_3$ is in the SP state at the highest fields~\cite{kubota2015successive,zhou2023possible}. 
As $\alpha$-RuCl$_3$ is believed to lie in the vicinity of the ferromagnetic-Kitaev limit~\cite{Matsuda2025}, the outstanding question is how the established KQSL response terminates and connects to the SP state.
Addressing this requires scrutinizing the broad field interval between the perturbative window and the high-field limit, where the collapse of the characteristic thermodynamic response would define the missing high-field part of the phase diagram.
Here, we explore this missing range by extending in-plane field-angle-resolved specific-heat measurements to higher fields up to 24\,T and complementing them with the in-plane magnetization. 

\section{Methods}
High-quality single crystals of $\alpha$-RuCl$_3$ were grown by the two-step sublimation technique~\cite{namba2024}.
The first step is a chemical vapor transport (CVT) purification, and the resulting crystals are then used as the source material for the second-step sublimation growth.
The specific heat sample has dimensions of $1950\times1080\times80\, \mu$m$^3$, which is the same ultraclean crystal as used in the previous study~\cite{namba2024,PhysRevB.111.184440} that benchmarked its sample quality.
For magnetization measurements, we used a separate piece from the same growth batch with dimensions of $1800\times1900\times350\,\mu$m$^3$.
In addition, a crystal grown by the vertical Bridgman method~\cite{kubota2015successive} was also used in this study, from which a particularly small piece with dimensions of $700\times500\times100\, \mu$m$^3$ was carefully selected. Its high quality was confirmed by magnetic susceptibility and specific heat measurements (see Appendix A).

The heat capacity was measured using a long-relaxation technique optimized for small-mass samples. 
Measurements were performed in a $^3$He refrigerator equipped with a two-axis rotator installed in a 25\,T cryogen-free superconducting magnet (25\,T-CSM) at the Institute for Materials Research (IMR), Tohoku University. A bare-chip resistive sensor (Cernox 1030BR) served as both thermometer and heater. The thermometer was calibrated in magnetic fields (up to 24\,T) against a reference calibrated thermometer. The crystal was mounted directly onto the bare chip using Apiezon grease to ensure good thermal contact.

High-field magnetization was measured by a standard induction method using a pickup coil with an inner diameter of 3\,mm in pulsed magnetic fields up to 50\,T. 
The pulsed fields were generated by a multilayer pulse magnet at the Institute for Solid State Physics, University of Tokyo. 
The half-cycle pulse had a duration of 4\,ms. 
A single crystal was placed inside the pickup coil and directly immersed in liquid $^4$He, which was cooled down to 1.35\,K by pumping.

\begin{figure*}[t]
    \includegraphics[width=0.9\linewidth]{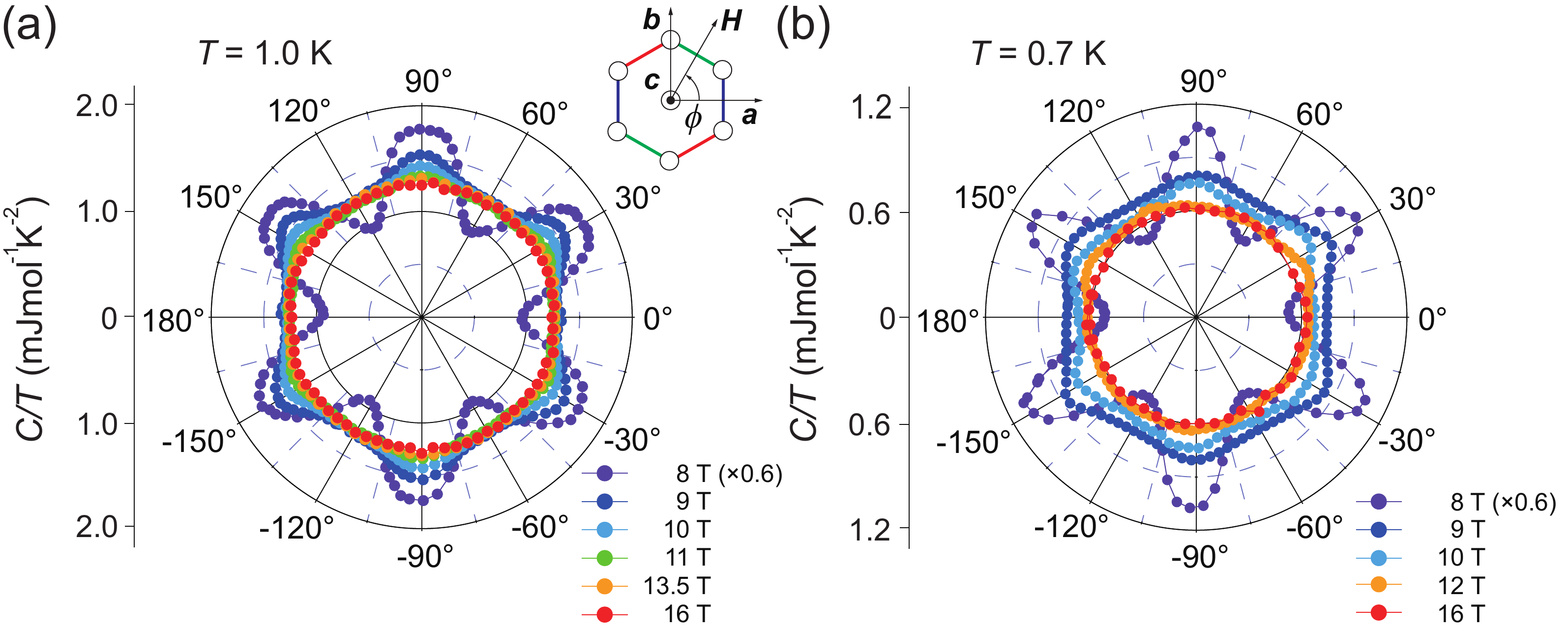}
    \caption{(a) Field-angle dependence of specific heat divided by temperature $C/T$ at $T=1$\,K as a function of the in-plane field angle $\phi$ for several magnetic fields, measured on a two-step sublimation crystal. Data in the range 8-11\,T are taken from the previous work~\cite{PhysRevB.111.184440}. The inset defines $\phi$ as the in-plane field-angle measured from the $a$ axis, which is perpendicular to a Ru-Ru bond direction of the honeycomb lattice.
(b) Field-angle dependence of  $C/T$ at $T=0.7$\,K as a function of $\phi$ for several magnetic fields, measured on a crystal grown by the Bridgman method.}
    \label{F1}
\end{figure*}

\section{Results}

First, we track the fate of the low-temperature in-plane angular response of the Majorana excitations by extending field-angle-resolved specific-heat measurements up to 16\,T. 
Figure\,\ref{F1}(a) shows the specific heat divided by temperature $C/T$ at $T=1$\,K for a crystal grown by the two-step sublimation method as a function of $\phi$, defined as the in-plane field-angle from the $a$ axis.
The data at 8 to 11\,T~\cite{PhysRevB.111.184440} show the characteristic angular modulation that has been discussed as a thermodynamic fingerprint of an anisotropic Majorana gap in the perturbative KQSL state.
With increasing fields, the amplitude of the six-fold oscillation in $C/T$ is progressively reduced, and eventually the response becomes nearly isotropic at 16\,T.
This pronounced suppression demonstrates that the characteristic angle-dependent low-energy response observed in the perturbative regime collapses at high fields.
Within the Majorana description, this behavior indicates that the low-energy thermodynamics can no longer be captured solely by the perturbatively induced, strongly angle-dependent Majorana gap $\Delta_{\text M}\propto|\kappa|$~\cite{KITAEV20062}. 
A natural interpretation is that, with increasing field, the Majorana and $Z_2$-flux sectors become increasingly mixed~\cite{PhysRevB.101.020414,PhysRevLett.119.127204,zhang2022theory}, introducing a more isotropic contribution to the low-energy spectrum.
As a result, the effectively gapless contribution for $\bm{H}\parallel\bm{b}$ is suppressed, and the low-energy response evolves toward an isotropically gapped spectrum.
Figure\,\ref{F1}(b) shows that the same overall trend is also observed in a crystal grown by the Bridgman method measured at $T=0.7$\,K. 
Notably, this Bridgman-grown sample has $T_{\text N}=7.6$\,K, higher than in conventional Bridgman-grown crystals~\cite{tanaka2020thermodynamic,imamura2024majorana} and comparable to that of the two-step sublimation crystals, indicating similarly high sample quality (see Appendix A). 
The common evolution seen in these two high-quality crystals clearly demonstrates that the six-fold oscillation is a robust signature of the perturbative KQSL regime, whereas the rounding of the angular response at higher fields reflects a general crossover out of the perturbative regime, independent of the growth method. 
In the Bridgman-grown sample, the angular dependence becomes weaker already from around 12\,T, reflecting the lower measurement temperature of $T=0.7$\,K.

\begin{figure*}[t]
    \includegraphics[width=1\linewidth]{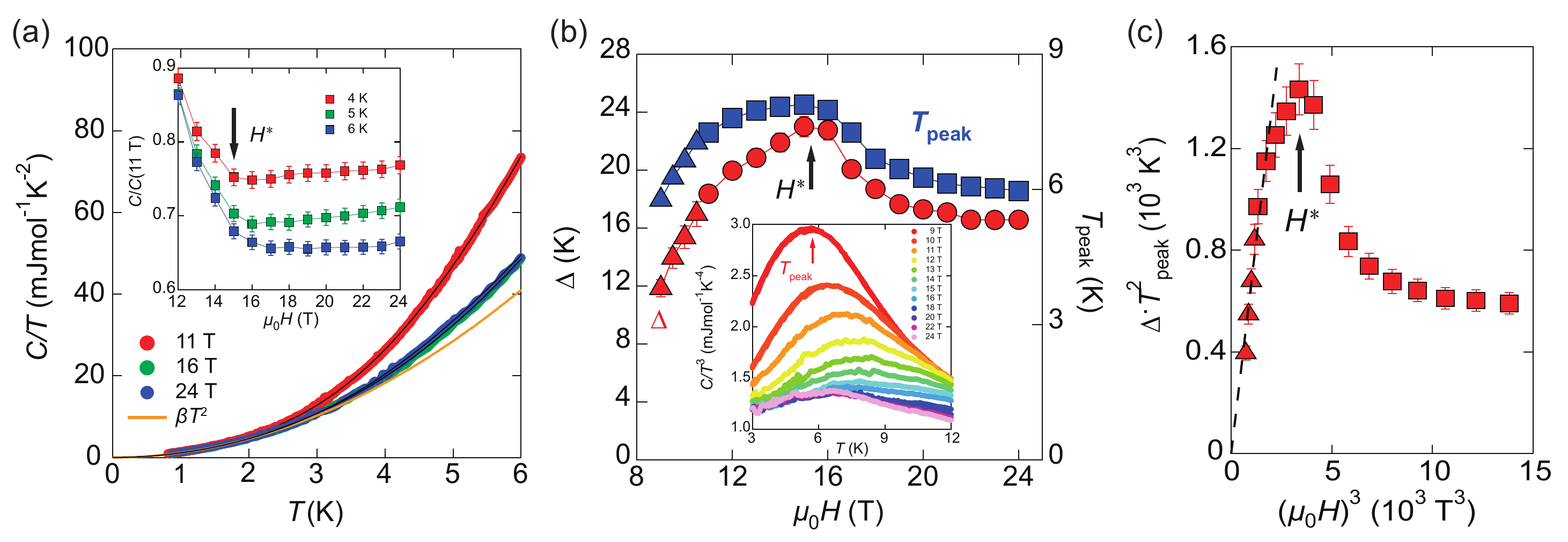}
    \caption{(a) Temperature dependence of the specific heat divided by temperature $C/T$ at 11\,T (red), 16\,T (green), and 24\,T (blue) for $\bm{H}\parallel\bm{a}$. 
The black solid lines show fits to the data below 6\,K. 
The orange line represents the phonon contribution $\beta T^2$ determined from the fit to the 24\,T data. 
The inset shows the field dependence of the specific heat $C$ at fixed temperatures of 4, 5, and 6\,K, normalized by the corresponding values at 11 T.
(b) Field dependence of the excitation gap $\Delta$ (red) and the peak temperature $T_{\text{peak}}$ (blue). Data below 11 T (triangles) are obtained by fitting the previously reported data using the same analysis procedure~\cite{PhysRevB.111.184440}. 
The inset shows the temperature dependence of $C/T^3$ at several fields up to 24\,T. 
The broad maximum in $C/T^3$ defines $T_{\text{peak}}$, as indicated by the arrow. Error bars in $\Delta$ represent the uncertainties obtained from the fits.
(c) Field evolution of $T_{\text{peak}}^2\Delta$ plotted against $H^3$. The black dashed line indicates the $H^3$ dependence in the lower-field regime, and the data deviate systematically from this relation on approaching $H^{*}$.}
    \label{F2}
\end{figure*}

Next, we focus on the temperature dependence of the specific heat to discuss how the excitation gap $\Delta$ and the associated thermodynamic energy scales evolve as the in-plane magnetic field is increased into the high-field regime.
Figure\,\ref{F2}(a) shows the temperature dependence of the measured $C/T$ at representative fields of 11, 16, and 24\,T for $\bm{H}\parallel\bm{a}$.
We fit the low-temperature $C/T$ data below 6\,K using $\beta T^2 + A{\text{exp}}(-\Delta/T)$, where the first term $\beta T^2$ accounts for the phonon contribution and the second term describes an activated low-energy excitation with an effective gap $\Delta$.
As shown by the black solid lines, this activated form reproduces the measured temperature dependence well over the fitting range. The orange line represents the phonon contribution determined from the fit to the 24 T data.
Overall, $C/T$ is suppressed as the magnetic field is increased. A closer comparison of the 16 and 24\,T curves, however, shows that they cross around 4\,K, suggesting a departure from a simple monotonic field evolution. 
To examine this behavior more directly, the inset of Fig.\,\ref{F2}(a) shows the field dependence of the unsubtracted specific heat at fixed temperatures of 4, 5, and 6\,K, normalized by the corresponding values at 11\,T. 
At all three temperatures, the specific heat decreases with increasing field up to around $\mu_0H^*\approx15$\,T and then increases again at higher fields. 
Thus, the nonmonotonic high-field evolution is directly visible in the raw specific-heat data without relying on phonon subtraction or fitting.
This fitting formula differs from our previous analysis, where the heat capacity was decomposed into three contributions from itinerant Majorana fermions, $Z_2$ flux excitations, and phonons~\cite{tanaka2020thermodynamic,PhysRevB.111.184440}. 
Even within the perturbative window the $Z_2$ flux contribution is small at low temperatures, and for $\bm{H}\parallel\bm{a}$ the itinerant Majorana spectrum is already gapped, so the present single-gap fit provides a qualitatively consistent approximation of the low-temperature behavior (see Appendix B).
We emphasize that $\Delta$ extracted in this way should be regarded as an effective thermodynamic gap associated with the low-energy sector described within the fractionalized-excitation framework, rather than the leading spin gap directly probed by spin-sensitive techniques such as ESR and NMR~\cite{PhysRevB.96.241107,baek2017evidence}, or the energy scales inferred indirectly from phonon scattering in longitudinal thermal conductivity~\cite{PhysRevB.102.235155}.
Figure\,\ref{F2}(b) summarizes the field dependence of the excitation gap $\Delta$ obtained from the fits and the characteristic peak temperature $T_{\text{peak}}$.
The excitation gap $\Delta$ obtained from the fitting increases with field at lower fields, but then exhibits a clear maximum at $H^*$, close to the field where the six-fold oscillation in $C/T$ disappears, and decreases upon further increasing the field.
The peak temperature $T_{\text{peak}}$ shows a similar nonmonotonic evolution. 
Here, $T_{\text{peak}}$ is defined as the temperature of the broad maximum in the directly measured $C/T^3$, as shown in the inset of Fig.\,\ref{F2}(b). 
The shift of this broad maximum with field provides a second indication, independent of the gap fitting, that the characteristic thermodynamic energy scale evolves nonmonotonically.
While the nonmonotonic behavior is already apparent in the unsubtracted specific-heat data, the validity of the phonon subtraction is supported by the field dependence of the fitted phonon coefficient $\beta$ (see Appendix B). 
With increasing field, $\beta$ converges monotonically to a nearly field-independent high-field value.
The high-field limiting value agrees well with that reported in a previous study, in which the lattice contribution of $\alpha$-RuCl$_3$ was independently estimated using both the specific heat of nonmagnetic RhCl$_3$ and $ab$ initio phonon calculations~\cite{PhysRevB.99.094415}. 
It is also consistent with the value obtained from our previous measurements on a different $\alpha$-RuCl$_3$ crystal~\cite{tanaka2020thermodynamic,Imamura2024defect}.
Furthermore, even when $\beta$ is fixed to this limiting value and the data are refitted over the entire field range, the extracted gap scale still exhibits the same nonmonotonic field dependence (see Appendix B).
These agreements support the physical validity of the adopted phonon background and show that the nonmonotonic behavior of  $\Delta$ and $T_{\text{peak}}$ is not an artifact of the phonon subtraction.

In the perturbative regime, $T_{\text{peak}}$ has been associated with the $Z_2$ flux energy scale and is expected to be proportional to the $Z_2$ flux gap $\Delta_{\text{flux}}$ ($T_{\text{peak}}\propto\Delta_{\text{flux}}$) from quantum Monte Carlo simulations~\cite{motome_nasu}. 
Combining this relation with the perturbative Majorana gap expression $\Delta_{\text M}\propto |h_xh_yh_z|/\Delta_{\text{flux}}^2$ yields the characteristic scaling $T_{\text{peak}}^2\Delta\propto H^3$.
As shown in Fig.\,\ref{F2}(c), the data approximately follow this $H^3$ dependence at lower fields but deviate systematically from it as the field approaches $H^{*}$. 
Together with the collapse of the six-fold angular oscillation, this deviation identifies $\mu_0H^*\approx15$\,T as the upper-field limit of the perturbative Kitaev response.
Notably, the observed decreasing trends of $\Delta$ and $T_{\text{peak}}$ with further increasing field are in qualitative agreement with theoretical calculations for the FM Kitaev model~\cite{hickey2019emergence,PhysRevB.100.144445}.
According to the theoretical studies, as the system approaches the SP phase, $\Delta$ and $T_{\text{peak}}$ are expected to pass through a minimum, and once the SP state is established they increase monotonically with field as the Zeeman gap grows~\cite{hickey2019emergence,PhysRevB.100.144445,PhysRevB.101.100408,okubo2025thermal}. 
However, such a monotonic increase of $\Delta$ and $T_{\rm peak}$ is not observed up to $\sim24$\,T. 
Their approach toward the expected minimum places the system already in the vicinity of the crossover toward the SP phase, while the absence of a monotonic increase indicates that it has not yet entered the trivial SP phase characterized solely by the Zeeman gap.
What we emphasize here is that the nonmonotonic field evolution of $\Delta$ and $T_{\text{peak}}$ cannot be explained without an intermediate regime, and therefore highlights the presence of an extended, nonperturbative KQSL regime at high fields above $H^*$ in $\alpha$-RuCl$_3$.

\begin{figure}[t]
    \includegraphics[width=1\linewidth]{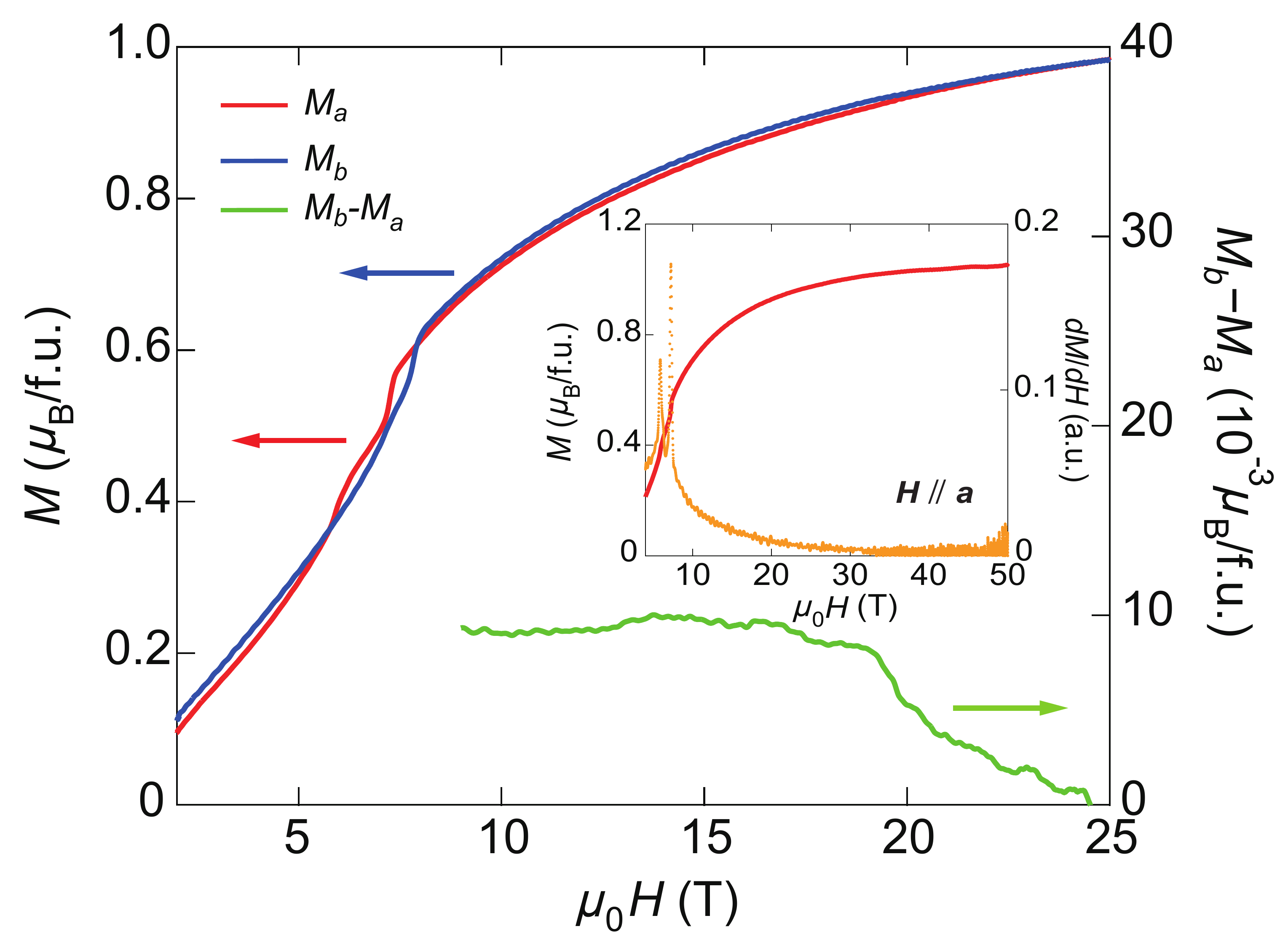}
    \caption{In-plane field dependence of magnetization for $\bm{H}\parallel\bm{a}$ ($M_a$, red) and  $\bm{H}\parallel\bm{b}$ ($M_b$, blue), together with the in-plane anisotropy $M_b-M_a$ at $T=1.35$\,K. The inset presents high-field magnetization $M_a$ (red) up to 50\,T for $\bm{H}\parallel\bm{a}$ and the corresponding differential susceptibility $dM/dH$ (orange).}
    \label{F3}
\end{figure}

We now turn to the magnetization as an independent thermodynamic probe of the high-field regime. In particular, we focus on the in-plane magnetization anisotropy. 
While the high-field magnetization of $\alpha$-RuCl$_3$ has been explored~\cite{kubota2015successive,zhou2023possible}, its in-plane anisotropy has not previously been tracked systematically into the high-field regime where the system approaches the SP phase.
Figure\,\ref{F3} summarizes the in-plane magnetization for $\bm{H}\parallel\bm{a}$ ($M_a$) and $\bm{H}\parallel\bm{b}$ ($M_b$), together with the anisotropy $M_b-M_a$. 
At low fields, $M_b$ exceeds $M_a$, and the anisotropy reverses sign in the vicinity of the transition field out of the ZZ phase, which is consistent with the field-angle-resolved magnetization measurements reported up to 7\,T~\cite{Balz2021}.
Upon entering the FIQD regime, $M_b$ becomes larger than $M_a$ again. At higher fields above $\sim18$\,T, $M_b-M_a$ starts to decrease, and eventually the in-plane anisotropy vanishes at $\sim 24$\,T.
This pronounced suppression establishes a characteristic high-field scale at which the anisotropic magnetic response of the intermediate-field regime is lost.
The inset extends the magnetization for $\bm{H}\parallel\bm{a}$ up to 50\,T and shows that $M_a$ approaches its saturation value at the highest fields, consistent with the approach to the spin-polarized limit at sufficiently high fields. 
Together with the absence of a sharp anomaly in $dM/dH$ beyond the ZZ transition around $\approx7$\,T, the suppression of $M_b-M_a$ is naturally interpreted as a crossover toward the SP phase. 
Notably, the vanished in-plane anisotropy at $\sim 24$\,T implies that the high-field in-plane response is effectively isotropic and therefore that a trivial anisotropic $g$ factor cannot account for the finite anisotropy observed in the intermediate-field regime. 
This supports that the anisotropy in the FIQD regime points to an intrinsic anisotropy of the low-energy response associated with the KQSL regime. 
The field range over which the magnetization anisotropy weakens and disappears is also consistent with the characteristic fields inferred from the angle-dependent specific heat and the extracted gap scale.
Taken together, these observations point to a distinct correlated intermediate regime in which the low-energy magnetic response remains qualitatively different from that of a conventional high-field SP phase.

\begin{figure}[t]
    \includegraphics[width=1\linewidth]{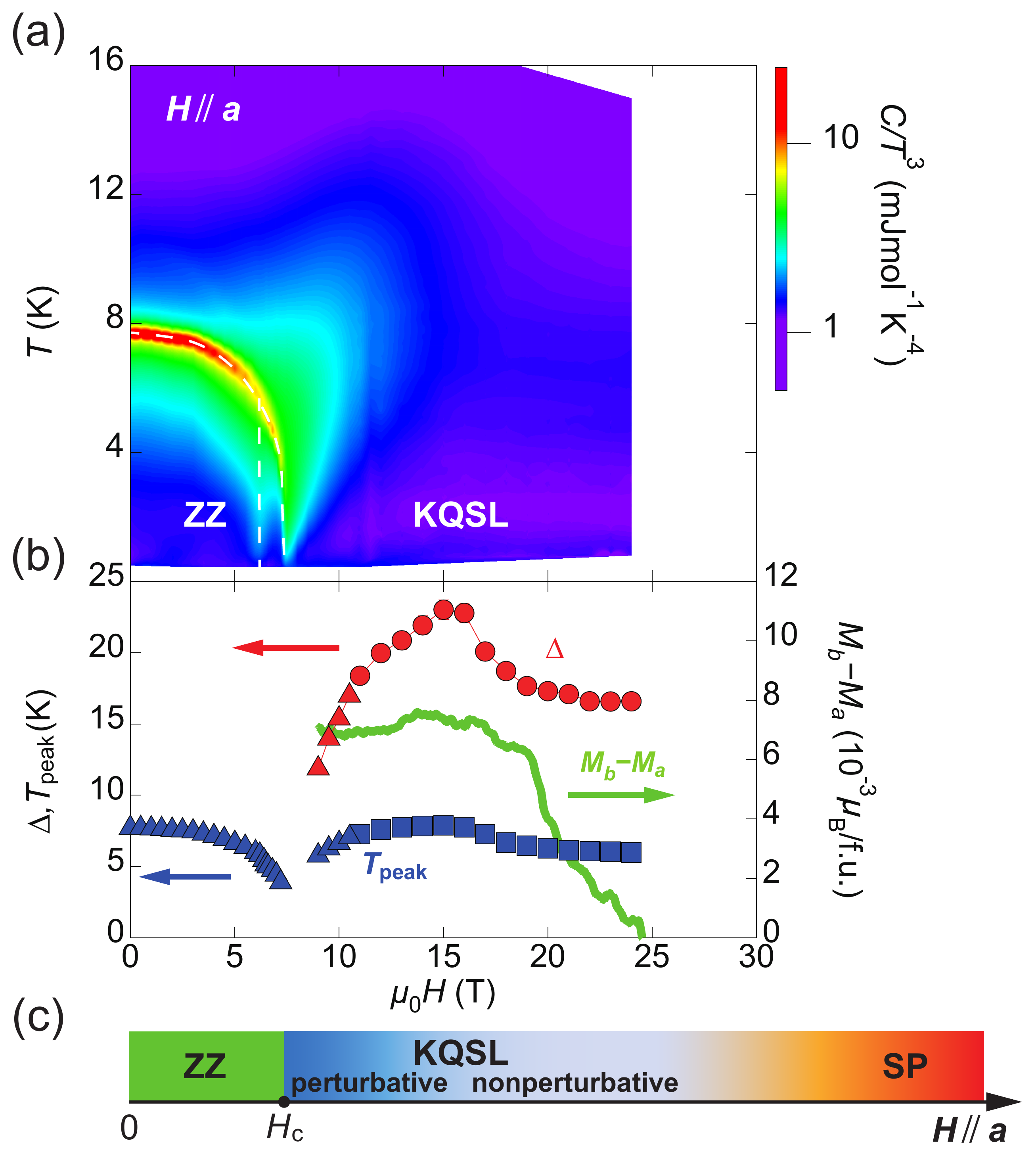}
    \caption{(a) Phase diagram in the field-temperature plane for $\bm{H}\parallel\bm{a}$. Superimposed color maps represent the magnitude of $C/T^3$.
(b) Field dependence of the excitation gap $\Delta$ (red), the peak temperature $T_{\text{peak}}$ (blue), and the in-plane magnetization anisotropy $M_b-M_a$ (green).
(c) Schematic high-field evolution for $\bm{H}\parallel\bm{a}$, summarizing the zigzag (ZZ) phase at low fields, the perturbative and nonperturbative KQSL regimes, and the crossover toward the spin-polarized (SP) phase at high fields.}
    \label{F4}
\end{figure}

\section{Discussion}

We finally turn to Fig.\,\ref{F4}, which presents the field-temperature phase diagram of $\alpha$-RuCl$_3$ for $\bm{H}\parallel\bm{a}$. 
Figure\,\ref{F4}(a) shows a color map of $C/T^3$. 
Since the phonon contribution scales as $T^3$, dividing by $T^3$ makes the phonon background approximately constant, so that the remaining field-temperature structure predominantly reflects non-phononic excitations. 
The data below 11\,T are obtained from previous work of the same ultraclean crystal~\cite{PhysRevB.111.184440} and are combined with the present high-field measurements to complete the phase diagram across the entire field range.
Figure\,\ref{F4}(b) summarizes the field evolution of $T_{\text{peak}}$, $\Delta$, and the magnetization anisotropy $M_b-M_a$.
At low fields, $T_{\text{peak}}$ corresponds to the magnetic transition temperature $T_{\text N}$ and decreases with increasing field, reflecting the suppression of the ZZ phase. 
Upon further increasing the field, the system enters the KQSL phase, where $T_{\text{peak}}$ increases again before being suppressed at higher fields as the system crosses over to the nonperturbative KQSL regime. 
A consistent trend is simultaneously observed in $\Delta$ and $M_b-M_a$, which evolve in the same field range.
Importantly, all three quantities exhibit correlated changes over a common field interval, providing a coherent set of magnetic field scales that demonstrate the progressive suppression of the KQSL regime before reaching the trivial SP phase.
Figure\,\ref{F4}(c) provides a schematic summary of the high-field evolution for $\bm{H}\parallel\bm{a}$. 
Guided by the magnetization data up to 50\,T, which indicate that the system reaches the SP limit at sufficiently high fields, the field-driven state can be grouped into four regimes: the ZZ phase, the perturbative KQSL regime, the newly found extended nonperturbative KQSL regime, and the SP phase. 
Although our thermodynamic measurements alone do not establish the topological nature of the high-field regime above $H^*$, its continuous evolution from the perturbative KQSL and its clear distinction from the conventional spin-polarized state motivate us to refer to this intermediate high-field regime as a nonperturbative KQSL.
It is striking that the trivial SP phase with increasing Zeeman gap is not realized at least up to 24\,T, where the magnetization reaches $\sim90$\% of saturation, implying the unexpectedly robust nature of the KQSL state against strong magnetic fields.
It should be noted that the transition from the ZZ phase to the KQSL regime is observed clearly at $\mu_0H_{\rm c}\approx7.5$\,T~\cite{PhysRevB.111.184440}, whereas a sharp transition from the KQSL phase to the SP phase is not resolved in the present magnetization data at finite temperatures. 
Although the KQSL-SP boundary involves a change of a topological invariant and thus corresponds to a phase transition strictly at zero temperature, a sharp transition between a two-dimensional spin-liquid regime and full spin polarization is well defined only in the zero temperature limit~\cite{hickey2019emergence,gordon2019theory,lee2020magnetic}.
At any finite temperature it is expected to appear as a smooth crossover. 
Moreover, even at zero temperature, the thermodynamic signatures near the boundary can be substantially broadened because sizable residual fluctuations persist near the boundary, including enhanced flux correlations~\cite{gordon2019theory,li2021identification}.
Therefore, the crossover behavior observed in the present thermodynamic properties is not in conflict with theoretical expectations~\cite{PhysRevB.101.100408,okubo2025thermal,PhysRevB.107.184418}.
At the same time, it is worth noting that in theoretical calculations based on more realistic spin models for $\alpha$-RuCl$_3$, which include non-Kitaev interactions such as Heisenberg and off-diagonal exchanges, stable intermediate phases are more commonly discussed for out-of-plane fields.
For in-plane fields, the ferromagnetic Kitaev-dominated regime is often found to evolve more directly toward spin polarization~\cite{gordon2019theory,li2021identification}. 
The robust high-field KQSL regime observed here therefore suggests that additional ingredients not fully captured in current modeling may play an essential role in $\alpha$-RuCl$_3$, such as interlayer couplings.

With the high-field thermodynamic landscape now delineated, the field window of the thermal Hall signal can be unambiguously located within the perturbative regime of the intermediate KQSL phase.
The most widely discussed thermal Hall signal in $\alpha$-RuCl$_3$ has been reported in an intermediate in-plane field range of roughly 9 to 13\,T~\cite{kasahara2018majorana,yokoi2020half,PhysRevB.106.L060410,PhysRevB.102.220404,bruin2022robustness,xing2025magnetothermal,Ongplanar,lefranccois2022evidence,Zhang2024}, well below $\mu_0H^*\approx15$\,T, overlapping with the perturbative KQSL window identified here.
This alignment strongly supports an interpretation in which the thermal Hall response originates from the field-induced topological Majorana band structure that is perturbatively generated in the KQSL regime. 
Accordingly, above $H^*$, the simple topological Majorana band picture underlying the quantized thermal Hall response would no longer apply, which could provide a natural explanation for the disappearance of the thermal Hall signal outside the perturbative regime.
By contrast, if the thermal Hall effect were instead dominated by topological magnons, the relevant topological bands would be most naturally defined in the SP phase~\cite{PhysRevLett.126.147201,PhysRevB.103.174402,PhysRevB.98.060404} or in the crossover regime on the way to spin polarization, where magnon quasiparticles become well defined~\cite{PhysRevB.107.184418}.
However, the high-field crossover regime identified in the present work lies above $H^*$, whereas the thermal Hall signal is confined to the lower-field perturbative KQSL regime. 
This clear separation of field scales is difficult to reconcile with a topological-magnon scenario associated with the crossover toward, or the establishment of, the SP phase. 
Instead, the one-to-one correspondence between the thermal Hall window and the perturbative thermodynamic signatures provides strong evidence that the thermal Hall response is tied to the field-induced Majorana band structure rather than to magnons emerging generically on approaching SP phase.

In conclusion, by combining field-angle-resolved specific heat with high-field magnetization anisotropy up to 24\,T, we have established a complete thermodynamic phase diagram of $\alpha$-RuCl$_3$ that maps the fate of the KQSL from the perturbative regime, through the nonperturbative regime, to the spin polarization. 
Our data identify a well-defined perturbative window below $\mu_0H^{*}\approx15$\,T, where the characteristic six-fold angular modulation and the  $H^3$ scaling are observed, and reveal that these hallmarks collapse above $H^*$. 
The excitation gap $\Delta$ and the peak temperature $T_{\rm peak}$ evolve nonmonotonically with fields, demonstrating that the route from the zigzag order to the SP state cannot be captured by a single direct transition but instead traverses perturbative and nonperturbative KQSL regimes before a broad crossover toward spin polarization. 
Strikingly, the KQSL signatures persist up to $\sim90$\% of magnetization saturation, revealing an unexpected robustness of the fractionalized state against strong magnetic fields.
These findings provide a bulk thermodynamic roadmap for how fractionalized phases evolve and ultimately yield to the spin polarization, opening a pathway toward exploring field-driven topological transitions and exotic excitations across Kitaev materials.

\section*{Appendix}

\subsection{Sample quality of the Bridgman crystal} 

\begin{figure}[t]
\includegraphics[width=0.8\linewidth]{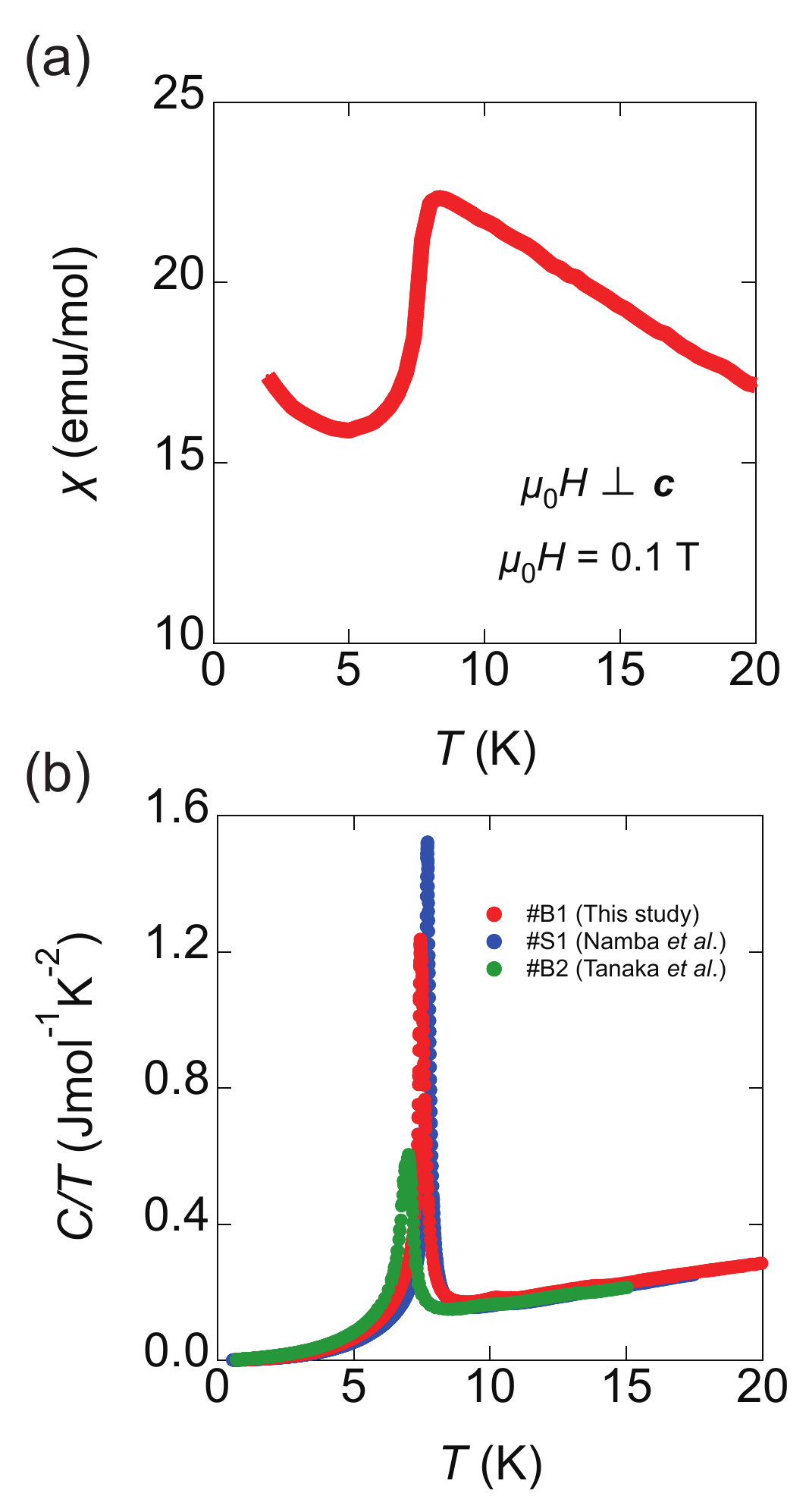}
\caption{(a) Temperature dependence of the magnetic susceptibility $\chi$ measured at $\mu_0H=0.1$\,T with ${\bm H}\perp{\bm c}$. (b) Temperature dependence of the specific heat divided by temperature $C/T$ for the present Bridgman-grown sample \#B1 (red), compared with the ultraclean two-step sublimation sample \#S1~\cite{namba2024} used in this study (blue) and the Bridgman-grown sample \#B2 used in the previous studies (green)~\cite{tanaka2020thermodynamic,imamura2024majorana}.}
\label{F5}
\end{figure}

To characterize the quality of the Bridgman-grown crystal \#B1 used in this study, we measured its low-field magnetic susceptibility and compared its specific heat with those of the samples used in the previous and present studies.
Figure\,\ref{F5}(a) shows the temperature dependence of the magnetic susceptibility measured at $\mu_0H=0.1$\,T with ${\bm H}\perp{\bm c}$. A sharp transition is observed at the transition temperature $T_{\text N}\approx7.6$\,K. 
This $T_{\text N}$ is higher than that of the Bridgman-grown sample \#B2 used in the previous studies ($T_{\text N}\approx7.0$\,K)~\cite{tanaka2020thermodynamic,imamura2024majorana}, and the data show no signatures of stacking faults, indicating minimal stacking faults in the present crystal \#B1.
Figure\,\ref{F5}(b) compares the specific heat divided by temperature $C/T$ at zero field of sample \#B1 with those of the ultraclean two-step sublimation sample \#S1~\cite{namba2024} used in this study and those of the Bridgman sample \#B2~\cite{tanaka2020thermodynamic,imamura2024majorana}. 
Sample \#B1 exhibits a sharp and clear jump in $C/T$ at $T_{\text N}$. 
Moreover, the $C/T$ value at $T_{\text N}$ is closer to that of the ultraclean sample \#S1 than that of sample \#B2, indicating that sample \#B1 is of similarly high quality as sample \#S1.
For the present measurements, we carefully selected a particularly small and clean piece from the grown crystals. The mass of sample \#B1 is 96\,$\mu$g, much smaller than that of sample \#B2 (680\,$\mu$g).

\begin{figure}[t]
\includegraphics[width=0.8\linewidth]{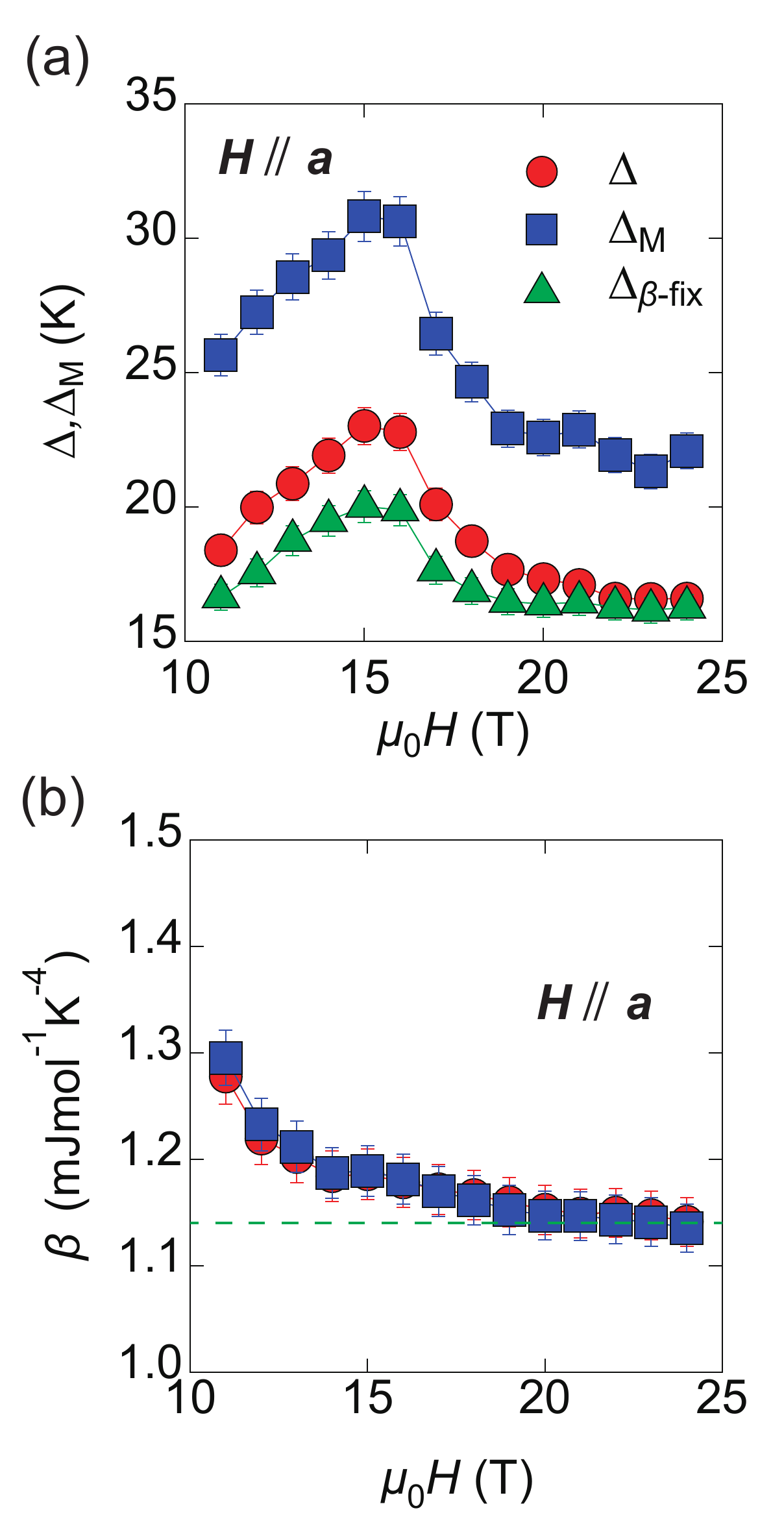}
\caption{(a) Field dependence of the gap obtained from three fitting procedures. Red circles show the effective gap $\Delta$ extracted using the activated form $C/T=\beta T^2 + A{\text{exp}}(-\Delta/T)$. Blue squares show the Majorana gap $\Delta_{\text M}$ obtained using the previous fitting form based on the two-dimensional gapped Majorana dispersion, $E=\sqrt{v^2|{\bm k}|^2+\Delta_{\text M}^2}$. Green triangles show $\Delta_{{\beta}{\text -}{\text{fix}}}$, obtained from the activated fit with the phonon coefficient $\beta$ fixed at its high-field limiting value indicated by the dashed line in (b). 
(b) Field dependence of the phonon coefficient $\beta$ obtained from the same two fitting procedures, with red and blue corresponding to the same fits as in (a), respectively. The green dashed line indicates the limiting value.}
\label{F6}
\end{figure}

\subsection{Robustness of the gap extraction against the fitting form and phonon background}

As described in the main text, the low-temperature $C/T$ data were fitted using $C/T=\beta T^2 + A{\text{exp}}(-\Delta/T)$, where the first term represents the phonon contribution and the second term represents a thermally activated contribution with an effective gap $\Delta$.
Here, we examine the robustness of the extracted field dependence of the gap by comparing three fitting procedures that differ in the functional form of the magnetic contribution and in the treatment of the phonon coefficient.
In the previous analysis, the low-temperature specific heat of itinerant Majorana fermions was described by the expression based on the two-dimensional gapped dispersion with velocity $v$, $E=\sqrt{v^2|{\bm k}|^2+\Delta_{\text{M}}^2}$, which was used to extract the Majorana gap $\Delta_{\rm M}$ in the perturbative regime.
The corresponding contribution to $C/T$ is given by $C_{\text{M}}(T;\Delta_{\text{M}})/T=\frac{\mathcal G(\infty)}{v^2}\left(1-\frac{\mathcal G(\Delta_{\text{M}}/T)}{\mathcal G(\infty)}\right)$, with $\mathcal G(y)=\int_0^y\frac{dx}{2\pi}\frac{x^3\text{e}^x}{(\text{e}^x+1)^2}$.
In addition to these two fitting forms, we performed a third analysis using the same activated form as in the main text while fixing $\beta$ to its high-field limiting value over the entire field range. The gap obtained from this procedure is denoted by $\Delta_{{\beta}{\text -}{\text{fix}}}$. 
This analysis provides a direct test of whether the field dependence of the fitted phonon background could artificially generate the observed evolution of the gap.

Figure\,\ref{F6}(a) compares the field dependence of the gap obtained from the three procedures.
The red circles show the effective gap $\Delta$ extracted using the activated form with $\beta$ treated as a fitting parameter, whereas the blue squares show the Majorana gap $\Delta_{\text M}$ obtained using the previous Majorana-based fitting form.
The green triangles show $\Delta_{{\beta}{\text -}{\text{fix}}}$, obtained from the activated fit with $\beta$ fixed to its high-field limiting value.
Although the absolute values of the extracted gaps depend on the fitting procedure, all three analyses yield the same qualitative field evolution: the gap initially increases with field, exhibits a pronounced nonmonotonic evolution near $H^{*}$, and decreases at higher fields. 
The agreement between $\Delta$ and $\Delta_{\text M}$ shows that this field dependence is insensitive to the detailed form assumed for the magnetic contribution. 
Moreover, the persistence of the same nonmonotonic behavior in $\Delta_{{\beta}{\text -}{\text{fix}}}$ demonstrates that it is not generated by allowing the phonon coefficient to vary independently at each field.
Figure\,\ref{F6}(b) shows the field dependence of the phonon coefficient $\beta$ obtained from the activated fit and the Majorana-based fit. 
The two fitting procedures yield nearly identical values of $\beta$ over the entire field range. 
In both analyses, $\beta$ decreases smoothly and monotonically with increasing field and converges toward a nearly field-independent high-field value, indicated by the green dashed line.
These results confirm that the fitting procedure does not affect the qualitative field evolution of the extracted gap.


\section*{Acknowledgments}
We thank E.-G.\ Moon and Y.\ Motome for fruitful discussions.
This work was performed at HFLSM under the GIMRT Program of the Institute for Materials Research, Tohoku University (Proposal No.\ 202412-HMKPA-0206).
This work was supported by a Grant-in-Aid for Scientific Research for Transformative Research Areas (A) ``Correlation Design Science'' (No.\ JP25H01248), Grants-in-Aid for Scientific Research (KAKENHI) (Nos.\ JP22H00105, JP24KJ0680, JP24H00937, JP23K26522, JP23K22447, JP24K21522, JP23KK0052, JP23H04014, JP23H04868, JP22H04933, JP25K00943 and JP26K21719) from Japan Society for the Promotion of Science (JSPS), CREST (No.\ JPMJCR19T5) and FOREST (No. JPMJFR236O) from Japan Science and Technology (JST), and Iketani Science and Technology Foundation.

\bibliography{Kitaev_ref_PRX}

\end{document}